# Polar Feature Based Deep Architectures for Automatic Modulation Classification Considering Channel Fading

Chieh-Fang Teng, Ching-Chun Liao, Chun-Hsiang Chen, An-Yeu (Andy) Wu, *Fellow*, *IEEE*
Graduate Institute of Electrical Engineering, National Taiwan University, Taipei, Taiwan
{jeff, cliao, johnny}@access.ee.ntu.edu.tw, andywu@ntu.edu.tw

*Abstract*—To develop intelligent receivers, automatic modulation classification (AMC) plays an important role for better spectrum utilization. The emerging deep learning (DL) technique has received much attention in AMC due to its superior performance in classifying data with deep structure. In this work, a novel polar-based deep learning architecture with channel compensation network (CCN) is proposed. Our test results show that learning features from polar domain ($r$-$\theta$) can improve recognition accuracy by 5% and reduce training overhead by 48%. Besides, the proposed CCN is also robust to channel fading, such as amplitude and phase offsets, and can improve the recognition accuracy by 14% under practical channel environments.

## I. INTRODUCTION

With the increasing volume of mobile devices and communication data rate, the radio spectrum is already congested. Hence, the future 5G wireless communication aims to achieve a quantum leap in data rate and spectral efficiency. To address aforementioned challenges with enhancement of Quality of Service (QoS), software-defined radio (SDR), cognitive radio (CR), and systems with adaptive modulations have been studies extensively [1]-[4]. All of them tend to develop ***intelligent modems***, aiming to fully utilize the radio spectrum. One new design is automatic modulation classification (AMC), which is the intermediate step between signal detection and demodulation [5]-[6]. In intelligent transceivers, based on the results of spectrum sensing, various modulation schemes are adopted to dynamically adjust transmission data rate and to meet the QoS requirement. At the receiver side, AMC is performed to blindly recognize the modulation types without prior knowledge of system parameters. Hence, with an effective AMC technique, the handshaking for exchanging *a priori* information can be removed, which leads to much fewer amount of transmitted data and lower transmission latency.

Recently, machine learning (ML) has been widely exploited to deal with the classification problem in AMC, for example, support vector machine, K-nearest neighbor, and genetic programming [7]-[9]. Without extracting features in hand-engineered, deep learning (DL) can automatically learn the high level features, and has received much attention due to its superior performance in classifying data with complex structure. Some existed works [10]-[13] have applied DL to the design of AMC. [11] and [12] introduced a one-dimensional Convolutional Neural Network (CNN) techniques for feature extraction. In [13], the raw received signals

This research work is financially supported in part by National Taiwan University and MediaTek Inc., Taiwan, under grants NTU-07HZA38001 and MTKC-2018-0167. The first two authors are also sponsored by MediaTek Ph.D. Fellowship program.

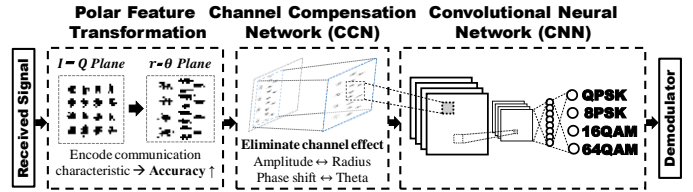

**Fig. 1.** Overview of proposed deep architecture for AMC.

(constellations) are converted into images for further classification. Due to CNN's superior performance in image recognition, [13] can achieve AMC very efficiently. However, two issues should be addressed:

1) *Format of input data*: Converting the received symbols into images and learning features from in-phase ($I$) and quadrature ($Q$) components of received symbols, in most of prior works, may cause accuracy degradation since the approaches lose the knowledge of communication systems.

2) *Channel fading*: only additive white Gaussian noise (AWGN) channel has been assumed in the simulation of previous works, which is unrealistic in practical implementations. The received constellation map must be distorted in amplitude and phase under channel impulse response which leads to severe performance degradation.

In this paper, by taking advantages of DL, we proposed a novel polar feature based deep architecture for AMC considering channel fading. Fig. 1 shows an overview of our proposed architecture, and our main contributions are summarized as follows:

1) Polar transformation technique is proposed to transform the received symbols from $I$-$Q$ to $r$-$\theta$ domain. Learning features from $r$-$\theta$ domain encodes specific knowledge of the communication system, and can improve the recognition accuracy by 5% and reduce training overhead by 48%.

2) A novel CNN architecture accompanied with channel compensation network (CCN) is designed to handle channel fading. Distorted received signals can be modified through CCN and achieve better performance.

The rest of this paper is organized as follows. Section II briefly reviews the signal model and prior works. Section III illustrates the proposed polar feature transformation technique and the deep architecture with CCN for AMC. The numerical experiments and analyses are shown in Section IV. Finally, Section V concludes this paper.

## II. BACKGROUND

### A. Signal Model

We assume that baseband in-phase ($I$) and quadrature ($Q$) components of $y(n)$ are extracted in a coherent, synchronous environment with single-tone signaling. In our approach, channel is treated as a flat fading in which frequency and phase offsets are added separately. The baseband sample of $y(n)$ after matched filtering can be expressed as:

$$y(n) = ae^{j(2\pi f_0 n + \theta_0)}s(n) + g(n), \quad (1)$$

where $a$ is an unknown amplitude factor, $f_0$ and $\theta_0$ are frequency offsets and unknown phase offsets respectively, $s(n)$ is transmitted symbol generated from one of selected modulations, and $g(n)$ is the complex Gaussian noise. Therefore, the modulation classification task is to blindly identify these modulation categories with the knowledge of the $L$-sample received symbol vector $\mathbf{y} = [y(1), y(2), \cdots, y(L)]^T$.

This paper considers four different modulation schemes, including quadrature phase-shift keying (QPSK), 8 phase-shift keying (8PSK), 16-quadrature amplitude modulation (16QAM), and 64-quadrature amplitude modulation (64QAM).

### B. Conventional Approach: Higher Order Statistic (HOS)

The two main traditional approaches of AMC are likelihood-based methods (LB) [14] and feature-based (FB) methods [15]. Despite of the optimal performance of LB, FB approaches are more widely adopted to yield sub-optimal performance with lower computational complexity. In [15], the higher order cumulants (HOCs) of fourth order are utilized for automatic modulation classification and the empirical cumulants of received symbol $\mathbf{y}$ are given by:

$$\hat{C}_{20} = \frac{1}{L}\sum_{n=1}^{L} y^2(n), \quad (2)$$

$$\hat{C}_{21} = \frac{1}{L}\sum_{n=1}^{L} |y(n)|^2, \quad (3)$$

$$\hat{C}_{40} = \frac{1}{L}\sum_{n=1}^{L} y^4(n) - 3\hat{C}_{20}^2, \quad (4)$$

$$\hat{C}_{41} = \frac{1}{L}\sum_{n=1}^{L} y^3(n)y^*(n) - 3\hat{C}_{20}\hat{C}_{21}, \quad (5)$$

$$\hat{C}_{42} = \frac{1}{L}\sum_{n=1}^{L} |y(n)|^4 - \left|\hat{C}_{20}\right|^2 - 2\hat{C}_{21}^2. \quad (6)$$

Thus, the lowest value of distance between empirical cumulants and theoretical values ($L \to \infty$) indicates the utilized modulation type.

### C. Deep Learning Approach: Image-based Convolutional Neural Network

Deep learning (DL), a promising branch of ML, is one of the fastest-growing and most breakthrough fields recently. Actually, as the traditional ML methods encounter bottlenecks such as the heavily relied experience-based knowledge and handcraft feature extraction, the most attractive advantage of DL is to learn features from data autonomously. In addition, according to the characteristics of input data, the architecture of DNN can be specific-modified. Convolutional neural network (CNN), for example, applies convolutional layers to the input signal in order to focus on the local connectivity and subtle features of data. Since most of patterns are much smaller than the whole image, applying CNN, a neuron does not have to see the whole image to discover the patterns. Additionally, the pooling layers, typically inserted after a convolutional layer, are also important for the CNN model. The most advantage of pooling layer is that it can reduce the dimensionality of the feature map and the number of parameters by doing the down-sampling operations. Meanwhile, it still keeps the most information of original feature maps.

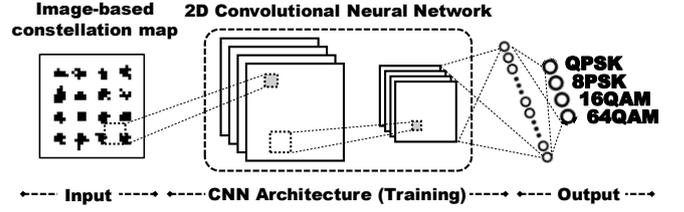

**Fig. 2.** Format of input data and architecture of prior work for AMC: Image-based 2-D CNN (I-Q Domain).

In [13], they exploited CNN model for the image-based modulation classification tasks. The received modulation symbols $y(n)$ as mentioned in (1) are complex points with $I$ and $Q$ components. Thus, the data conversion is necessary to bridge the gap between complex symbols and images. [13] mapped the received symbols into scatter points on the complex plane. The constellation diagram in $I$-$Q$ plane is shown in Fig. 3(b). Both real and imaginary axes range are fixed from -3.5 to 3.5 and the image resolution are set to be 227×227 as the input setting of AlexNet model [16]. The image-based input data and architecture are shown in Fig. 2.

## III. PROPOSED DEEP ARCHITECTURE FOR AMC

### A. Polar Feature Transformation

According to the method used by [13], they directly mapped the complex symbols to a constellation image and have a good performance through powerful CNN. Although the method is simple and useful, learning from images in $I$-$Q$ plane loses the knowledge of the communication systems. In addition, it is apparent that the constellation of QPSK can be seen as a sub-picture of 8PSK. Likewise, 16QAM also has correlation with 64QAM. The nested constellations sharing points in $I$-$Q$ based images result in high misclassification between them.

Therefore, we propose a polar feature transformation to map the complex symbols to images in polar coordinate. Learning features from $r$-$\theta$ domain encodes specific knowledge of the communication systems. Taking spectrum analysis for example, we certainly can input time-series data into learning process. The training result might lead to a Fast Fourier Transform (FFT) with lot of noise term. However, if we apply existed expert knowledge, data can undergo FFT before training. The training result would perform better and require less training time as shown in Section IV.B.

In this paper, leveraging existed expert knowledge in communication, we can transform $I$-$Q$ domain into $r$-$\theta$ domain before learning. Besides, learning in $r$-$\theta$ domain have another advantage. We can deal with channel fading directly in $r$-$\theta$ domain. The constellation map would be distorted through the channel impulse response. The distortion contains amplitude scaling and phase rotation. Amplitude can be mapped into $r$ axis and phase can be represented by $\theta$ axis. In $r$-$\theta$ domain, it is easier to learn the modified parameter ($\Delta r$, $\Delta \theta$) to eliminate the channel fading. This will be introduced in the Section III.B.

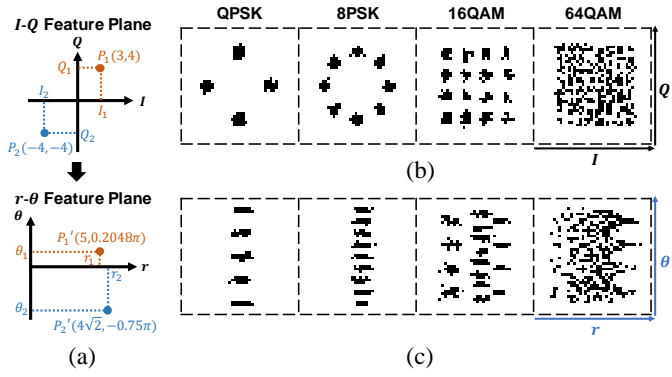

**Fig. 3.** Input data in constellation diagrams of four modulation categories with SNR=20dB: (a) illustration of the process of polar feature transformation; (b) *I-Q* plane; (c) *r-θ* plane.

To construct the relation between $I$ and $Q$ components, we associate with polar coordinate which replaces the $I$-$Q$ axis with $r$-$\theta$ axis, as illustrated in Fig. 3(a). The transformation function can be expressed as:

$$r[n] = radius[n] = \sqrt{I[n]^2 + Q[n]^2}, \quad (7)$$

$$\theta[n] = theta[n] = arctan(\frac{Q[n]}{I[n]}), \; n = 0,1,...,L-1, \quad (8)$$

where $L$ is the symbol length and $I$, $Q$ represent in-phase and quadrature components respectively. The polar feature based images of selected modulation categories are depicted in Fig. 3(c) where the horizontal and vertical axis represent radius and theta respectively. The radius range is set $[0, 3]$, theta range is $[-\pi, \pi]$ and the image resolution is 36×36.

### B. Deep Architecture with Channel Compensation Network

The architecture of our system is illustrated in Fig. 4. It can be mainly divided into two parts, CNN architecture and channel compensation network (CCN). The architecture of CNN model is constructed by four convolutional layers and three dense layers, as shown in Fig. 4(b). The nonlinear activation function, Rectified Linear Units (ReLUs), among each layer is defined as:

$$f_{ReLU}(x) = \max\{0, x\}. \quad (9)$$

The activation functions can result in a nonlinear system which is helpful for extracting more complex features than a high-order linear system such as matched filter. In addition, the regularization technique of "dropout", which avoids updating the weights of part nodes, is also utilized to reduce overfitting and force the nodes be more independent than usual. Finally, the loss function, categorical cross-entropy, is set to compute the error gradient between true labels $m_i$ and predict labels $\widehat{m}_i$ and defined as:

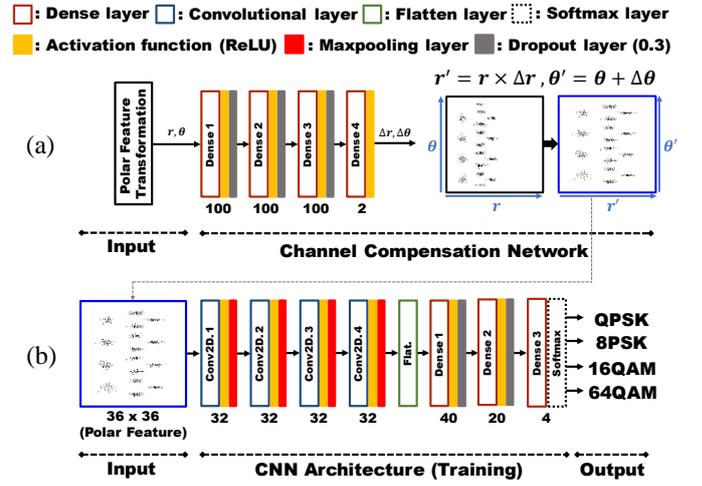

**Fig. 4.** The detailed overview of deep architecture with channel compensation network (CCN): (a) architecture of CCN; (b) CNN architecture.

$$\mathcal{L}(m_i, \widehat{m}_i) = m_i log(\widehat{m}_i) + (1 - m_i)log(1 - \widehat{m}_i). \quad (10)$$

Only additive white Gaussian noise (AWGN) channel has been assumed in the simulation of prior works. However, in realistic communication systems, the received constellation map must be distorted in amplitude and phase under channel impulse response, which make the recovery of transmitted signal more challenging. Definitely, the channel fading also have huge impact on the accuracy of modulation classification. In [13], they achieve good performance of accuracy with the consideration of channel is ideal and only with the impact of AWGN. Therefore, we want to put our system to a more realistic environment to examine the impacts on our accuracy and try to eliminate the channel fading simultaneously in $r$-$\theta$ domain.

The common channel effects on constellation maps are difference of received power and phase shift that mean the pattern on images may scale and rotate with a random number. To reduce the impact of scaling and phase shift on the performance, we try to find out the inverse channel response. In [17], the team of Google DeepMind proposed a spatial transformer network to learn an appropriate transformation for the input picture. This is particularly useful for the application of image recognition where the input images may suffer from warping, rotate, and scale. With the spatial transformer, the model can be more spatially invariant to the input data.

Inspired by this work, we design a channel compensation network to learn the inverse channel parameters and connect to the CNN model as illustrated in Fig. 4(a). There are four dense layers in the network. The learned parameters of inverse channel response, $\Delta r$ and $\Delta \theta$, are utilized to eliminate the channel effects and the reconstructed signals can be expressed as below:

$$r' = r \times \Delta r, \; \theta' = \theta + \Delta \theta. \quad (11)$$

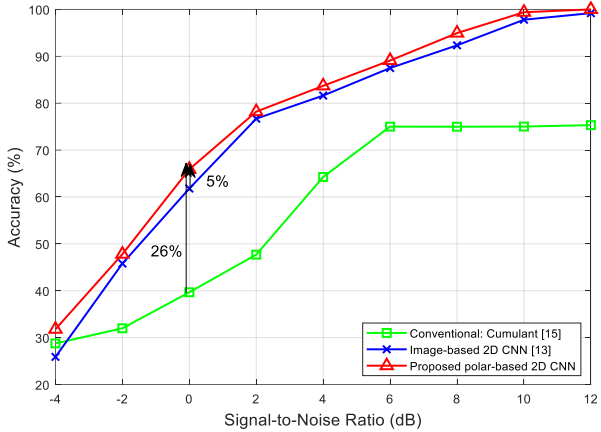

**Fig. 5.** Recognition accuracy of the proposed polar feature based deep architecture versus SNR.

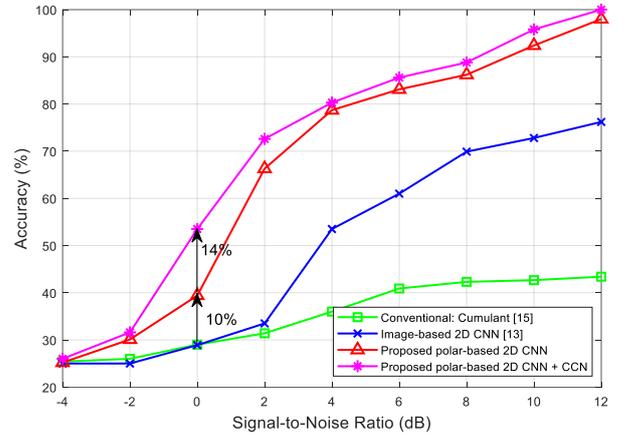

**Fig. 6.** Recognition accuracy under channel fading and the accuracy compensated by CCN.

where $r$ and $\theta$ are directly transformed from the received signal $y(n)$ by the method proposed in Section III.A. Then, the reconstructed signals are mapped to $r$-$\theta$ plane as input data for the concatenated CNN model. This neural network is just like the equalizer in communication systems which is important for the compensation of channel distortion and is helpful to reduce the transmission bit-error-rate (BER).

## IV. EXPERIMENTAL RESULTS AND ANALYSIS

The simulation setup is summarized in Table I.

### A. Deep Architecture w/o Channel Fading

The first experiment is to simulate the classification accuracy of proposed method under different noise interference. The value of the SNR of the AWGN changes from -4dB to 12dB. The accuracy is averaged for four categories of modulation techniques. Besides, in this section, we assume that the channel is ideal only with AWGN. The simulation results are shown in Fig. 5. We compare the deep architecture we proposed in $r$-$\theta$ domain with fourth order cumulants approach and image-based method in $I$-$Q$ domain. The accuracy of our proposed method improve 5% and 26% recognition accuracy than image-based and cumulants approach when SNR equals to 0dB.

### B. Run Time of Different AMC Approach

Furthermore, the training and inference time of different AMC approaches are listed in Table II. From the simulation results, the proposed approach reduces the training overhead about 48% compared to image-based approach which indicates that learning in $r$-$\theta$ domain not only has better performance but also has faster convergence speed. Besides, the inference time is short enough for real time applications.

### C. Deep Architecture w/ CCN under Channel Fading

To make the system more practical, we consider the channel with variance of received power scaling and phase shift that are common phenomenon in OFDM system. Simultaneously, we utilize the trained CCN to learn the inverse channel response and compensate for the channel distortion.

**TABLE I**. Simulation parameters.

| Modulation category | QPSK, 8PSK, 16QAM, 64QAM |
|---|---|
| Training data/modulation | 5000 images |
| Testing data/modulation | 1000 images |
| Symbol length ($L$) | 1000 |
| Signal to Noise Ratio (SNR) | -4, -2, 0, 2, 4, 6, 8, 10, 12 |
| Training and testing environment | Deep learning library of Keras running on top of TensorFlow with NVIDIA GTX 980 Ti GPU |

**TABLE II**. Run time.

| Different AMC Approach | Training Time with GPU (s) | Inference Time with CPU (s) |
|---|---|---|
| **Image-based 2D CNN [13]** | 52.9556 | 9.31e-04 |
| **Proposed polar-based 2D CNN** | 27.5152 | 9.31e-04 |

All the simulation setups are kept the same as given in the previous experiment. The simulation results are shown in Fig. 6. When considering to the channel fading, the performance degrades significantly. However, our proposed polar feature is still better than image-based method which demonstrates that our method is more tolerant to the channel distortion. Besides, the CCN, which can compensate the channel distortion, is also helpful for improving the recognition accuracy by 14% under a practical environment.

## V. CONCLUSIONS

In this paper, we present a novel polar feature based deep architecture with channel compensation network for AMC. It can learn from $r$-$\theta$ domain to achieve better recognition accuracy. Moreover, the proposed CCN can compensate for the distorted signal in realistic channel. Our proposed design can be used as an enabling technique for realizing intelligent receiver.

REFERENCES

[1] Jondral, Friedrich K. "Software-defined radio: basics and evolution to cognitive radio," EURASIP journal on wireless communications and networking 2005.3 (2005): 275-283.

[2] K. E. Nolan, L. Doyle, P. Mackenzie, D. O'Mahony, "Modulation scheme classification for 4G software radio wireless networks," Proc. IASTED. 2002.

[3] Haykin, Simon. "Cognitive radio: brain-empowered wireless communications," IEEE journal on selected areas in communications 23.2 (2005): 201-220.

[4] H. Sun, A. Nallanathan, C.-X. Wang, Y. Chen, "Wideband spectrum sensing for cognitive radio networks: A survey," IEEE Wireless Commun., vol. 20, no. 2, pp. 74-81, Apr. 2013.

[5] O. A. Dobre, A. Abdi, Y. Bar-Ness, W. Su, "Survey of automatic modulation classification techniques: Classical approaches and new trends," IET Commun., vol. 1, no. 2, pp. 137-156, Apr. 2007.

[6] Z. Zhu, A. K. Nandi (2015). "Automatic Modulation Classification: Principles, Algorithms and Applications," London : Wiley

[7] Cheol-Sun Park, Won Jang, Sun-Phil Nah. and Dae Young Kim, "Automatic Modulation Recognition using Support Vector Machine in Software Radio Applications," in Proc. 9th IEEE ICACT, Feb. 2007, pp. 9-12

[8] X. Zhou, Y. Wu, B. Yang, "Signal Classification Method Based on Support Vector Machine and High-Order Cumulants," Sicintific research, Wireless Sensor Networks, No. 2, pp. 48-52, Nov. 2009.

[9] M. W. Aslam, Z. Zhu, A. K. Nandi, "Automatic Modulation Classification Using Combination of Genetic Programming and KNN," IEEE Transactions on wireless communications, vol.11, no.8, pp.2742-2750, Aug. 2012.

[10] G. J. Mendis, J. Wei, A. Madanayake, "Deep learning-based automated modulation classification for cognitive radio," 2016 IEEE International Conference on Communication Systems (ICCS), pp. 1-6, 2016.

[11] O'Shea, Timothy J., Johnathan Corgan, and T. Charles Clancy, "Convolutional radio modulation recognition networks," International Conference on Engineering Applications of Neural Networks. Springer International Publishing, 2016.

[12] Nathan E. West and Timothy J. O'Shea, "Deep architectures for modulation recognition," In 2017 IEEE International Conference on Dynamic Spectrum Access Networks, March 2017.

[13] Peng, Shengliang, *et al.*, "Modulation classification using convolutional Neural Network based deep learning model," in 2017 26th Wireless and Optical Communication Conference (WOCC), Newark, NJ, USA, 2017.

[14] J.A. Sills, "Maximum-likelihood modulation classification for PSK/QAM," in Proc. IEEE MILCOM, Atlantic City, NJ, Oct. 31-Nov. 3 1999.

[15] A. Swami and B. M. Sadler, "Hierarchical digital modulation classification using cumulants," IEEE Trans. Commun., vol. 48, no. 3, pp.416–429, Mar. 2000.

[16] A. Krizhevsky, I. Sutskever and G. Hinton, "ImageNet Classification with Deep Convolutional Neural Networks," in Advances in Neural Information Processing Systems 25, pp. 1097-1105, 2012.

[17] Jaderberg, Max, Karen Simonyan, and Andrew Zisserman. "Spatial transformer networks," Advances in Neural Information Processing Systems. 2015.